\documentclass[a4paper,11pt]{article}
\pdfoutput=1

\usepackage{jinstpub} 

\usepackage{lineno,hyperref}
\usepackage[utf8]{inputenc} 
\usepackage{nth}
\usepackage{amssymb,amsmath}
\usepackage{color}
\usepackage{caption}
\usepackage{subfig}
\DeclareMathOperator{\sgn}{sgn}

\usepackage{natbib}
\newcommand{\figw}{0.7\textwidth}

\title{Harmonizing discovery thresholds and reporting two-sided confidence intervals: a modified  Feldman \& Cousins method}
\author[]{K. D. Morå}

\affiliation[]{Oskar Klein Centre, Department of Physics, Stockholm University\\Albanova University Center, SE-10691 Stockholm, Sweden}

\emailAdd{knut.mora@fysik.su.se}
\abstract{
When searching for new physics effects, collaborations will often wish to publish upper limits and intervals with a lower confidence level than the threshold they would set to claim an excess or a discovery. In this paper a modification to the Feldman-Cousins method is proposed that allows for a transition from one-sided upper confidence limits for null results and a two-sided confidence intervals for non-null results at any given specified threshold chosen to define the observation of a signal, while maintaining exact coverage.
}

\begin{document}
\maketitle
\flushbottom
\section{Introduction}
Many physics experiments, in particular searches for new physics, look for very low event rates where the asymptotic methods of constructing frequentist confidence intervals do not work. Confidence intervals are required to have coverage;  $1-\alpha$-confidence level intervals should contain the true value a fraction $1-\alpha$ of repeated experiments. However, the actual coverage of a statistical method may vary with the true signal properties. For example, an asymptotic $0.68$ confidence-level interval for a counting experiment observing $n$ events , $[n-\sqrt{n},n+\sqrt{n}]$, will cover the true expectation value $\mu$ asymptotically as $\mu\to\infty$, but may cover as little as $0.55$ and as much as $1$ depending on $\mu$.  

\begin{figure}
    \centering
    \includegraphics[width=\figw]{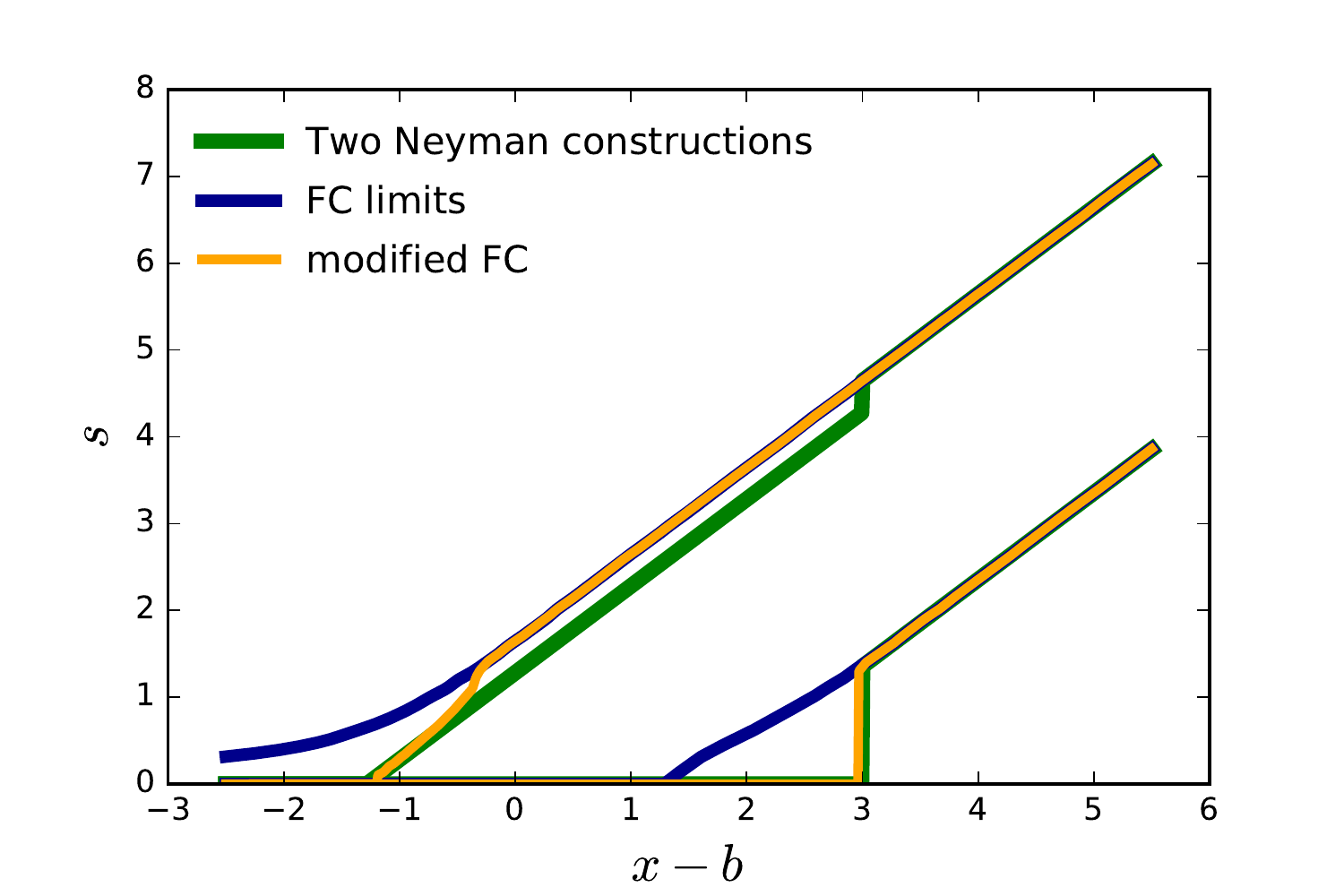}
    \caption{Illustrations of three constructions of upper and lower limits for a Gaussian observable $x$, with known background $b$. The green lines show the upper and lower limit as function of $x-b$ for an experiment that switches between setting a $90\%$ upper limits for discovery significances below $3\sigma$, and uses a two-sided interval above. The blue lines shows the FC upper and lower limits. Orange lines shows the modified FC method, that like the FC method provides coverage for all true signals, but switches between a one- and two-sided limit when the threshold significance of $3\sigma$ is reached. This leads the upper limit for this construction to approach the one-sided limit construction for low $x+b$.}
    \label{fig:limits_gaussian}
\end{figure}

A method that provides confidence intervals with exact coverage is known since 1937 as the Neyman construction \cite{Neyman:1937uhy}. The Neyman construction initially consists of constructing a confidence belt for each possible true value of parameter of interest $s$:
\begin{equation}
\label{eq:confidencebelt}
1- \alpha = \int_a^b f(x|s) dx
\end{equation}
where $f(x|s)$ is the probability density function for the observed parameter $x$, which may depend on $s$, and $[a,b]$ denote the limits of the confidence belt.
The confidence interval on $s$ can then be found by constructing $a(s)$ and $b(s)$, which will express the upper and lower range in which $x$ would fall $1-\alpha$ of the time if the true parameter of interest is $s$. Inverting these functions yields the Neyman construction limits for an observation $x$: 
\begin{equation}
	 [a^{-1}(x), b^{-1}(x) ] 
\label{eq:NeymanLimit}
\end{equation}
The condition for the confidence belt provided in equation \ref{eq:confidencebelt} is not unique, and the limits of the confidence belt have to be set by a boundary condition. This condition has traditionally consisted in either the desire to set upper or lower limits (for example in absence of a signal) or in reporting (symmetric or asymmetric) two-sided intervals in case of a measurement of a physical parameter.
In some cases, such as in searches for a new particle, experiments may wish to report upper limits if they do not observe a discovery significance exceeding a set threshold, and a two-sided confidence interval otherwise. 
However, Feldman \& Cousins \cite{Feldman:1997qc} noted that the fact that switching between Neyman constructions based on the experimental outcome may lead to under-coverage, even if the individual constructions provide coverage. 
The suggested remedy (hereafter referred to as the "FC method") is to construct confidence intervals by a single Neyman construction that provides both upper limits and two-sided intervals, depending on the experimental result. The FC confidence belt, reviewed in section~\ref{sec:FC}, uses the log-likelihood ratio to decide which regions of observable space to include first. Figure~\ref{fig:limits_gaussian} shows the upper and lower limits for a Gaussian observable $x$ with known background $b$ and standard deviation $1$ for the FC construction in blue, and an experiment that switches to two-sided intervals from upper limits if the discovery significance exceeds $3\sigma$ in green. Since this shift moves only the upper limit line for, for example $s=2$, this approach will under-cover for this signal. The modification suggested in this paper, which is constructed to maintain coverage is shown in orange, and may be interpreted as a coverage-conserving interpolation between a one- and two-sided Neyman construction.

%
%
%

Conventionally, upper limits are reported with confidence levels of less than 95\%, and two-sided intervals  are presented only in the case of discovery or at least some reasonably significant indication. The (one-sided) p-value of an indication is usually much smaller  than the 5 \% or more $\alpha$ implied by the $1-\alpha$ confidence interval. While statistically presenting a two-sided interval and not claiming a discovery does not pose a problem (the fact that the confidence interval excludes the non existence of a signal at some confidence level should not be confused with a discovery claim), in practice experimenters are reluctant to present a two sided limit even if the FC method provides it.
A common remedy is to report only the upper edge of the interval provided.
This leads to a signal-dependent over-coverage, or, equivalently, some confidence intervals or upper limits could be more constraining without violating coverage. 	
In this paper, we suggest a modification of the FC method that will provide two-sided intervals only at a desired discovery threshold, while still providing a unified confidence interval calculation method and improving the coverage. The paper is organized as follows: in section \ref{sec:FC} we review briefly the FC method, and the procedure for assessing the existence of an excess or a discovery. In section \ref{sec:mFC} we introduce the modified version of the FC method, and we illustrate the method with line-search example in section~\ref{sec:linesearch}.

\section{Feldman-Cousins construction}
\label{sec:FC}
For an experiment where one measures some data $\vec{x}$ with a probability distribution $f(\vec{x} | s)$ that depends on a parameter $s$, the likelihood is given as $\mathcal{L}(s) = f(\vec{x}|s)$. The method proposed by Feldman and Cousins uses the log-ratio $R$ between the likelihood given $s$, and the $s$ that minimizes the likelihood, $\hat{s}$:
\begin{equation}
\label{eq:likelihoodratio}
R(\theta) = 2\cdot\log\left[\mathcal{L}(\hat{s})/\mathcal{L}(s)\right]
\end{equation}
to decide which $\vec{x}$ to include. 
Either constructing the confidence belt from Equation\ \ref{eq:confidencebelt} with the constraint that the $\vec{x}$ with the lowest $R(s)$ are included first, or constructing the confidence belt directly in the $R(s)$ parameter:
\begin{equation}
\label{eq:confidencebelt_FC}
1- \alpha = \int_0^{R_\mathrm{max,\alpha}(s)} f(R|s) dR
\end{equation}
for each value of $s$ will yield the FC construction. 
The confidence interval, whether one- or two-sided will be the region where $R(s)<R_\mathrm{max,\alpha}(s)$. 
Note that the threshold likelihood ratio $R_\mathrm{max,\alpha}(s)$ also depends on the parameter of interest. 
In the case that an experiment is looking to constrain a parameter $s$ that has a null-hypothesis and lower bound, $s_0$, the method has to give confidence intervals that do not contain $s_0$ in $\alpha$ of the cases.
For example, in searches for the production cross-section of an unknown particle, $\alpha$ of confidence intervals will exclude the no-signal null-hypothesis.


The log-likelihood ratio $R(s_0)$ is typically also used to assess discovery significance with respect to the null-hypothesis $s=s_0$. 
The p-value of $R(s_\mathrm{result})$ under the null-hypothesis is:
\begin{equation}
\label{eq:pval_FC}
p_\mathrm{result}(R_\mathrm{result}(s_0)) = \int_{R_\mathrm{result}(s_0)}^\infty f(R|s_0) dR
\end{equation}
This may also be inverted to yield discovery thresholds; $p_{result}^{-1}(\alpha)$ is the discovery threshold for an $\alpha$ excess. Note that this equation shows that at the null-hypothesis $s=s_0$, the FC threshold for inclusion in the confidence interval, $R_\mathrm{max,\alpha}(s_0)$ implies a p-value of $\alpha$, and that a confidence interval that does not include $s_0$ implies a p-value below $\alpha$.
Typical confidence intervals for upper limits, and thus the FC construction are $\alpha=0.1,0.05,0.01$. Using the FC method consistently will report two-sided intervals at those same thresholds. However, a conventional discovery threshold in particle physics is $5\sigma$, or $p=3\cdot10^{-7}$, and experiments may not wish to publish measurements of excesses lower than, say, $3\sigma$, or $p=1.3\cdot10^{-3}$. A pragmatic solution to this is to only report the upper edge of the confidence interval as an upper limit until the discovery significance has exceeded the required discovery threshold. This will lead to over-coverage, as one extends a confidence interval constructed to cover with an $1-\alpha$ frequency. 

\section{Modified Feldman-Cousins}
\label{sec:mFC}
The aim of the modified method is to provide a construction with a desired discovery significance threshold, different than what the confidence level would imply, in addition to maintaining the constant coverage of the pure FC construction. 
In figure~\ref{fig:limits_gaussian}, this modification is indicated with an orange line, showing that the modification does not change the FC construction at higher signals, while approaching the one-sided Neyman construction upper limit for low signals. In this illustration, the modification does not reach the median signal-free result of $x-b=0$, but for higher discovery thresholds, such as $4\sigma$, even the median upper limit will be affected by the modification, as shown in the coverage plots in figure~\ref{fig:coverages} for the example in section~\ref{sec:linesearch}. 

\begin{figure}
    \centering
    \includegraphics[width=\figw]{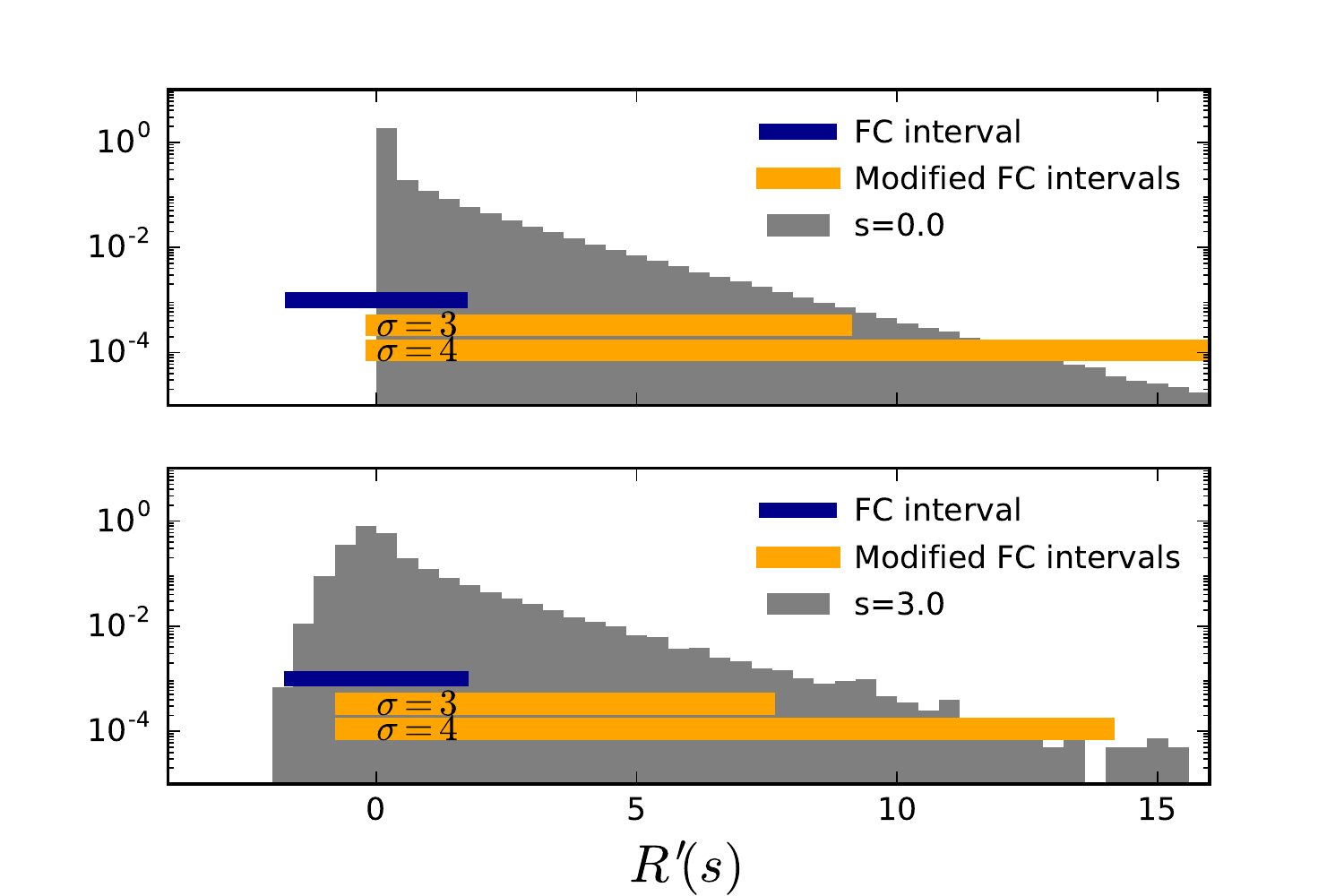}
    \caption{Histograms of $R'(s)$ computed for toy-Monte Carlo simulations for $0$ and $3$ expected signal events in the upper and lower panel, according to the example in section~\ref{sec:linesearch}. The best-fit signal rate $\hat{s}$ is constrained to be non-negative in the signal model and fit. The sharp boundary at $R'(s)=0$ in the upper parameter will be seen in all cases where  the constraint $0\leq\hat{s}$ is applied to the best-fit. Blue and orange bands show the $90\%$ confidence band for the FC method and the modified FC method, respectively, with the latter shown both for a $3\sigma$ and $4\sigma$ discovery threshold.}
    \label{fig:powerhist}
\end{figure}

We wish to include all results where the discovery significance is less than the reporting threshold in our Neyman confidence belt, while maintaining coverage for all signals. To accomplish this, we will treat upwards and downwards fluctuations separately, and include all upwards fluctuations that do not rise to the reporting threshold in our band. This will require constricting the confidence band for downwards fluctuations to conserve coverage. 
To distinguish between upwards and downwards fluctuations, the proposed modification to the FC method uses an idea very similar as that which was used by the \textsc{atlas} Higgs search~\cite{Aad:2012an}; where the ordering ratio $R$ is multiplied by the sign of $\hat{s}-s$:
\begin{equation}
\label{eq:fcminus_rsign}
R'(s) = \sgn(\hat{s}-s)\cdot R(s) \end{equation}
This separates the cases where the data prefers a lower and higher signal than the tested hypothesis. Close to a boundary, say a requirement that $s_0\leq s$, $R'(s_0)$ can only be non-negative, and for slightly larger $s$, the distribution of $R'(s)$ will still be asymmetric between upwards and downwards fluctuations. 
The switch of sign of $R'(s)$ occurs as $\hat{s}$ approaches $s$, which is also where $R(s)$ approaches $0$.   
Examples of the distributions of $R'(s)$ for the line-search detailed in the next section are shown in figure~\ref{fig:powerhist}, including a blue line indicating the ranges of $R'(s)$ corresponding to a $90\%$ confidence level FC interval. Orange lines show that the modified FC method band shifts to include more of the positive $R'(s)$ in order to avoid excluding excesses below the discovery thresholds indicated. 

We will construct the edge of the confidence belt corresponding to upwards fluctuations, $R^+(s)$,  first. We denote the confidence level of the interval $1-\alpha$, and the p-value threshold for reporting a two-sided excess $\gamma$.
To ensure that our confidence intervals exclude the null hypothesis case when the discovery significance exceeds the reporting threshold, $R^+(s_0)$ must correspond to the discovery threshold $R_\mathrm{max,\gamma}(s_0)$ defined in equation~\ref{eq:pval_FC}. At large signals, we wish $R^+(s)$ to approach the FC edge $R_\mathrm{max,\alpha}(s)$. We accomplish this by interpolating between the two thresholds:
\begin{equation}
\label{eq:fcminus_r_plus}
R'(s)^{+} = w(s)\cdot R_\mathrm{max,\gamma}(s_0) + (1-w(s))\cdot R_\mathrm{max,\alpha}(s)
\end{equation}
where $w(s)$ is a weighting function that monotonically decreases from $1$ at $s=s_0$ to 0 as $s$ increases. The FC threshold function,  $R_\mathrm{max,\alpha}(s)$ is defined by equation \ref{eq:confidencebelt_FC}. 
The freedom to choose $w(s)$ reflects the original freedom in the Neyman construction. However, we wish the confidence band to rapidly approach the FC band with increasing $s$. In some simple cases, such as a single Gaussian distributed variable with known standard deviation, an observation with a discovery significance exactly equal to the threshold will have an $R'(s)$-curve that exactly divides observations below or above the discovery threshold, and the $R^+(s)$ curve may be constructed as the maximum of this curve and the FC threshold. This corresponds to the vertical line at $x-b=3$ in figure~\ref{fig:limits_gaussian}.

\begin{figure}
    \centering
    \includegraphics[width=\figw]{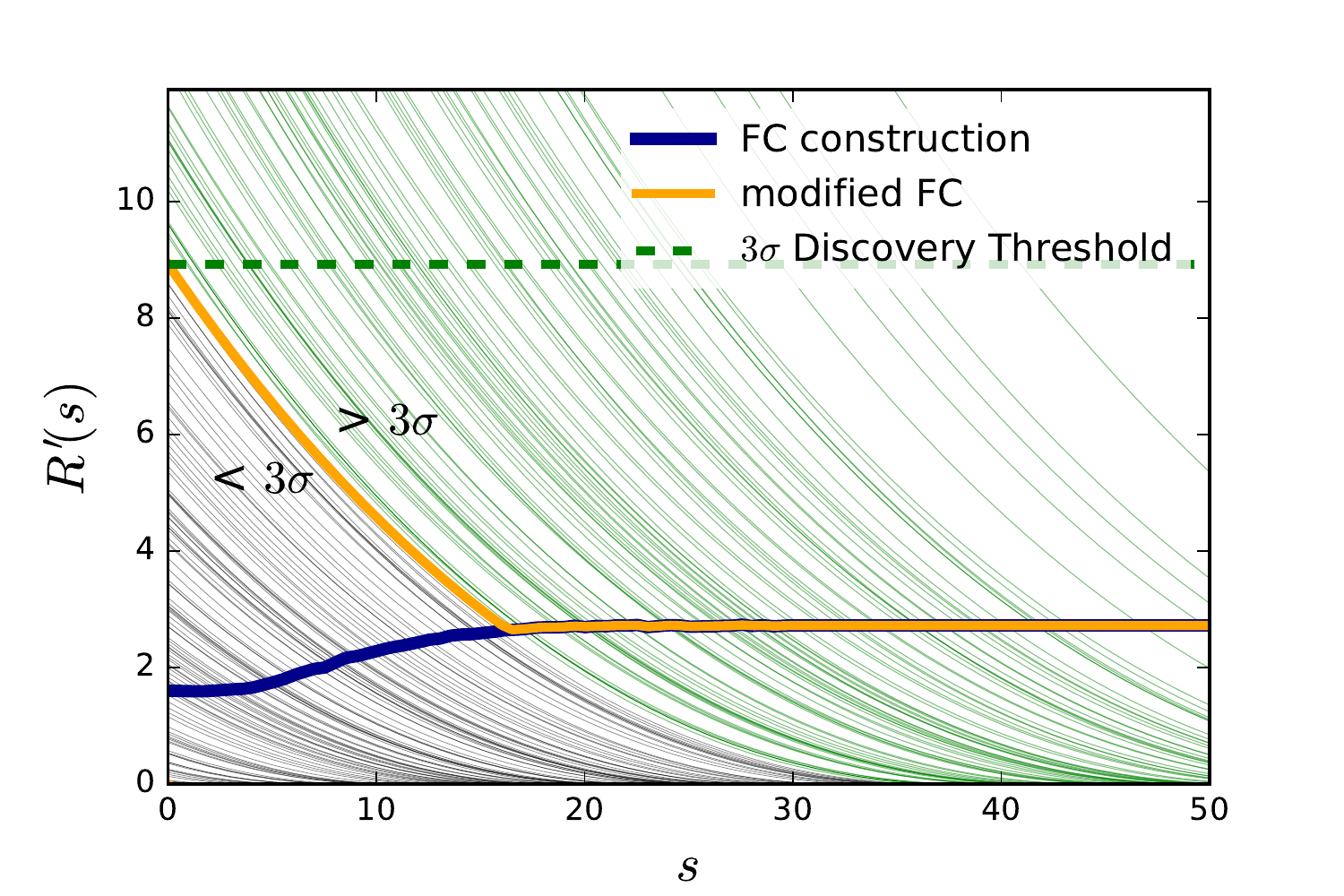}
    \caption{Construction of the modified FC threshold function, using curves of $R'(s)$ for multiple toy-Monte Carlo realizations, colored according to whether they exceed the $3\sigma$ discovery significance threshold (green) or not (gray). The thick blue curve shows the threshold corresponding to a $90\%$ FC construction, while the orange curve, showing $R^+(s)$, is constructed to be equal to the discovery threshold, indicated with a dashed green line, at 0 signal, and to be greater or equal to both the likelihood-ratio curves with discovery significance less than $3\sigma$ and the FC threshold.}
    \label{fig:envelope}
\end{figure}

In the more general case, a toy Monte Carlo approach can be used, by generating toy data sets with a range of true signal sizes, and selecting all observations whose discovery significance does not exceed the threshold. We construct the likelihood ratio curves $R'_i(s)$ for all these observations, labelled with index $i$, and compute the maximum value at each signal, $R_\mathrm{envelope}(s) = \sup_i (R'_i(s))$. Finally, we set the threshold $R^+(s)$ to be the greatest of $R_\mathrm{envelope}(s)$ and $R_\mathrm{max,\alpha}(s)$. This construction is shown in figure~\ref{fig:envelope}. 

The lower edge of the interval, $R^-(s)$ is then defined so that for any signal, $1-\alpha$ of $R'(s)$s are contained between $R^-(s)$ and $R^+(s)$:
\begin{equation}
\label{eq:fcminus_r_minus}
1-\alpha = \int_{R^{-}(s)}^{R^+(s)} f(R'|s) dR'
\end{equation}
At $s=s_0$, the above equation would indicate a coverage of $1-\alpha$. However, at the border of the domain for $\alpha$, the distribution of $R'(s_0)$ will be peaked towards $0$, as shown in the lower panel of figure~\ref{fig:powerhist} , and by defining the confidence interval lower threshold $R^-(s_0)=-\epsilon$, where $\epsilon$ is an infinitesimally small negative number such that no $R'(s_0)$ are lower than $R^-(s_0)$, the coverage at the boundary can be arranged to be $\gamma$.

Confidence intervals are constructed as in the FC case as intersections between $R'(s)$ and $R^+(s)$ for lower limits, and $R'(s)=R^-(s)$ for upper limits. 
For increasing discovery thresholds, the discovery belt must move to include higher positive values of $R'(s)$, and correspondingly increase $R^-(s)$ as well. Close to $s_0$, where the shift is the highest, $R^-(s)$ will approach the Neyman construction boundary for an upper-limit-only construction, which can provide empty confidence intervals for strong but finite downwards fluctuations of the background. The FC method yields higher limits in the downwards fluctuation regime, as illustrated in figure~\ref{fig:limits_gaussian} for the Gaussian example, with the upper limit approaching zero asymptotically when the downwards fluctuation approaches negative infinity. 
Some experiments setting upper limits have adopted the CLs method~\cite{cls}, which penalizes the p-value to yield a signal-dependent over-coverage at low signal-background discrimination approaching $1$ for signals approaching $s_0$. Others have used a power-constraint~\cite{Cowan:2011an}, where upper limits are not placed below a signal where the experiment has a certain discovery power. Direct detection experiments using the two-sided FC method~\cite{Aprile:2018dbl,PhysRevLett.118.021303}, have applied a power-constraint, corresponding to a $-1\sigma$ downwards fluctuation of upper limit. The coverage properties of the power-constraint applied to the modified FC confidence intervals has a simpler form than the CLs method, with the coverage being $1-\beta(s)$ below the critical discovery power, where $\beta(s)$ is the discovery power, and $0.9$ above. The threshold discovery power constraint cannot be found from first principles, similarly to the lower limit reporting threshold, but must be chosen by the experiment.
In the next section, an example using toy Monte Carlo computations is shown, illustrating the method, and showing that the required coverage is achieved. 

\section{Example Application for Line-Search}
\label{sec:linesearch}
As an example, we consider an experiment that observes events with energies $E_i$, and searches for a Gaussian signal line with a certain mass in the presence of a power-law background. The probability distribution function $f(E)$ has the form: 
\begin{align}
    f(E|s) &=& \frac{s}{(s+b)}f_s(E)+\frac{b}{s+b}f_b(E) \\
    f_s(E) &\equiv& \frac{1}{\sqrt{2\pi\sigma}}e^{\frac{(E-m)^2}{-2\sigma^2}}\\
    f_b(E) &\equiv& E^{\Gamma}\cdot\left[\int_{E_0}^{E_1}E^\Gamma\mathrm{d}E\right]^{-1}
\end{align}
Here, $s$ is the signal expectation value, and $m,\sigma=m/5$ are the signal mass and energy uncertainty characterizing the signal distribution $f_s(E)$. The background power-law $f_b(E)$ has expectation value $b=1000$, and power-law index $\Gamma=-2$. Both distributions are normalized between $E_0$ and $E_1=100 E_0$. 
The signal expectation value is not allowed to be negative in the fit, $0\leq\hat{s}$. 
In this example, the nuisance parameters affecting the distribution shapes, and the background expectation are fixed. 
For most experiments, the likelihood will include a number of nuisance parameters. In that case, the ordering parameter $R$ may be based on the profiled likelihood instead, yielding the profile construction~\cite{PDG}, where coverage is not ensured by construction, but must be investigated. 
The toy-Monte Carlo methods used to construct the confidence belt and to investigate coverage properties are identical to the ones used for the profile construction.

\begin{figure}
    \centering
    \includegraphics[width=\figw]{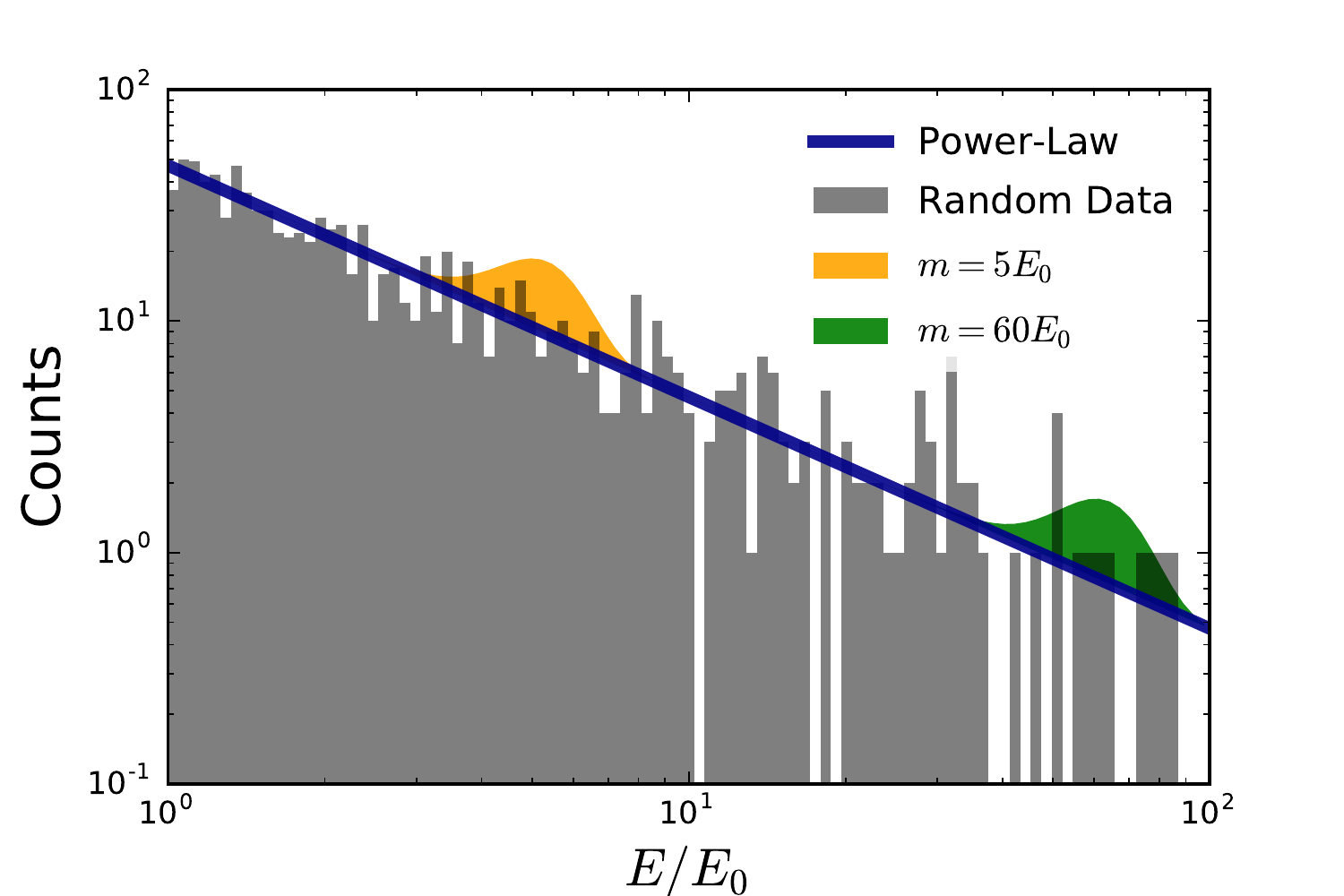}
    \caption{Background  distribution (blue line), and signal distributions for $m=5E_0$ and $m=60E_0$ (orange and green), together with a histogram showing an example data-set drawn from the background-only distribution.}
    \label{fig:powerspectrum}
\end{figure}

The extended un-binned likelihood for the observation of $N$ energies with values $E_i$ can be written: 
\begin{align}
    \mathcal{L}(s|N,\vec{E}) = \mathrm{Pois}(N|s+b)\cdot \prod_{i=1}^{N}\left[f(E_i|s)\right]
\end{align}
Here, $\mathrm{Pois}(N|s+b)$ is the Poisson probability to observe $N$ events given an expectation value of $s+b$. The distribution of the signed ordering parameter $R'(s)$, defined from equations~\ref{eq:likelihoodratio} and~\ref{eq:fcminus_rsign}, is shown in figure~\ref{fig:powerhist}, for a signal mass $m=5 E_0$, and for two different true signal expectations $s=0,3$. As an example of using toy Monte-Carlo methods to determine the interpolation function $w(s)$, Figure~\ref{fig:envelope} shows multiple curves of $R'(s)$ from toyMC simulations with true signals ranging from $0$ to $40$, divided by whether the discovery significance, assessed with $R'(0)$, is above or below a $3\sigma$ discovery threshold. The weighting function $w(s)$ is chosen so that the modified FC threshold is equal or greater than all the $R'(s)$ curves for all signals, until the $R'(s)^+$-curve, in orange, meets the FC belt in blue. The $90\%$ confidence belts derived from this construction are also shown in figure~\ref{fig:powerhist} as bands for $3$ and $4\sigma$ discovery thresholds.

\begin{figure}
    \centering
    \subfloat[$m=60$]{\includegraphics[width=0.5\linewidth]{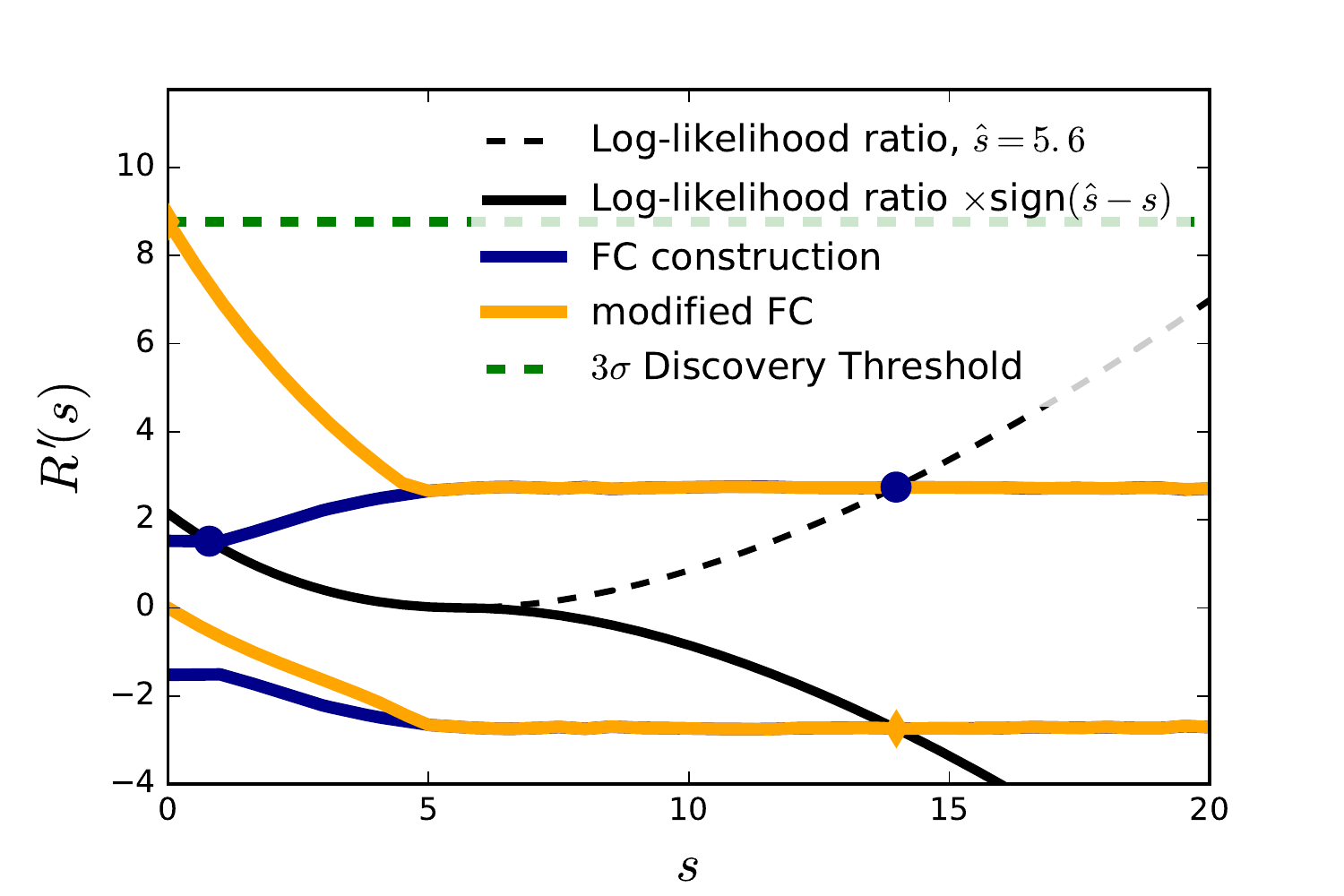}}
    \subfloat[$m=5$]{\includegraphics[width=0.5\linewidth]{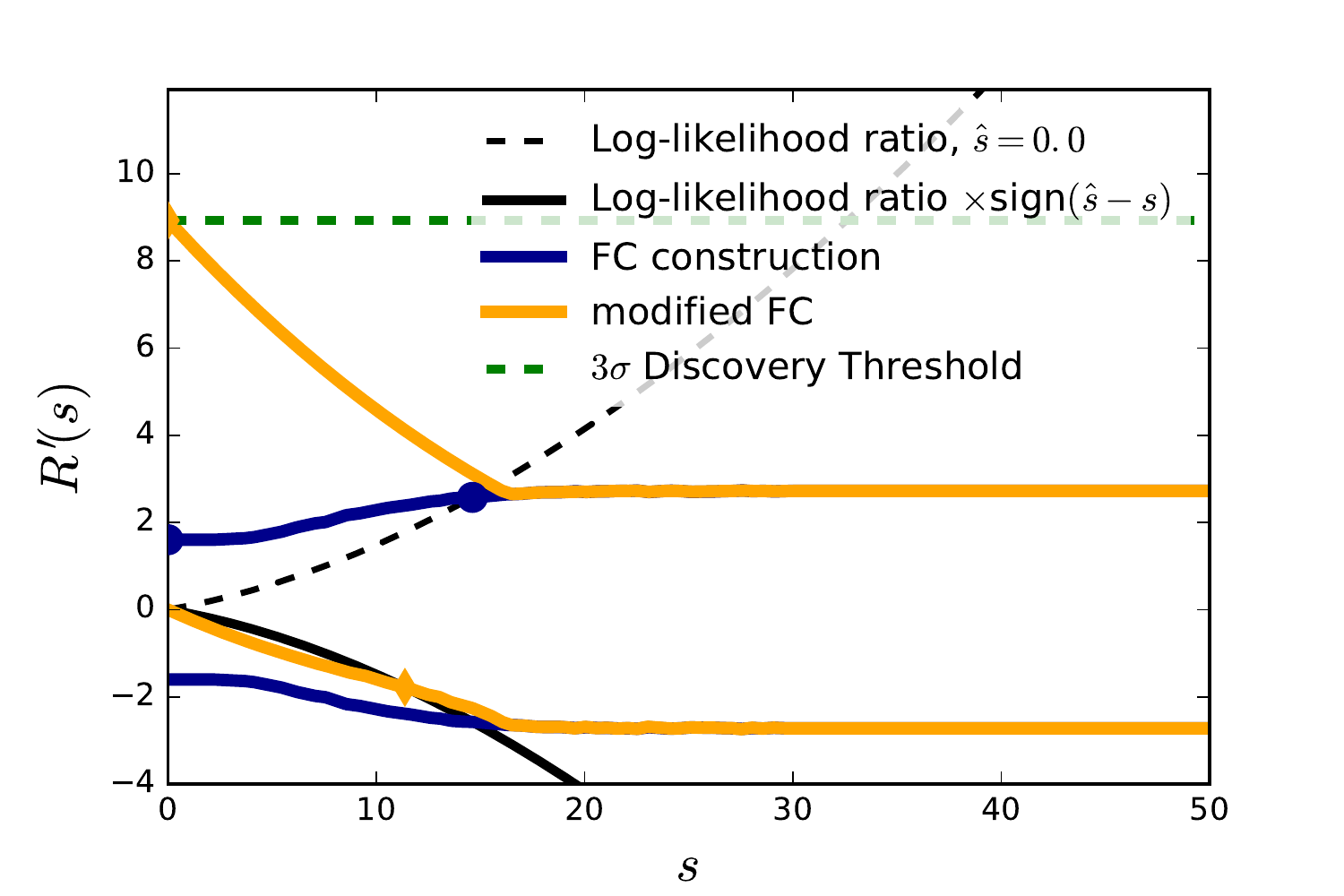}}
    \caption{Illustration of confidence interval constructions using the FC (blue bands) and the modified FC (orange bands). Intersections between the $R'(s)$-curve, in black and the thresholds define the upper and lower limits of the interval. For comparison, the FC construction is shown with the dashed black line, with confidence interval boundaries marked by blue dots. }
    \label{fig:ciconstruction}
\end{figure}

The confidence interval construction for the power-law example is shown in figure~\ref{fig:ciconstruction}, showing both cases where an excess with a p-value below $10\%$ gives a two-sided interval, and a case where both constructions yields upper limits. The confidence interval consists of the signal range where $R'(s)$ is contained between the $R^+(s)$ and $R^-(s)$ curves. 

\begin{figure}
    \centering
    \subfloat[$m=5$, $3\sigma$ threshold]{\includegraphics[width=0.5\linewidth]{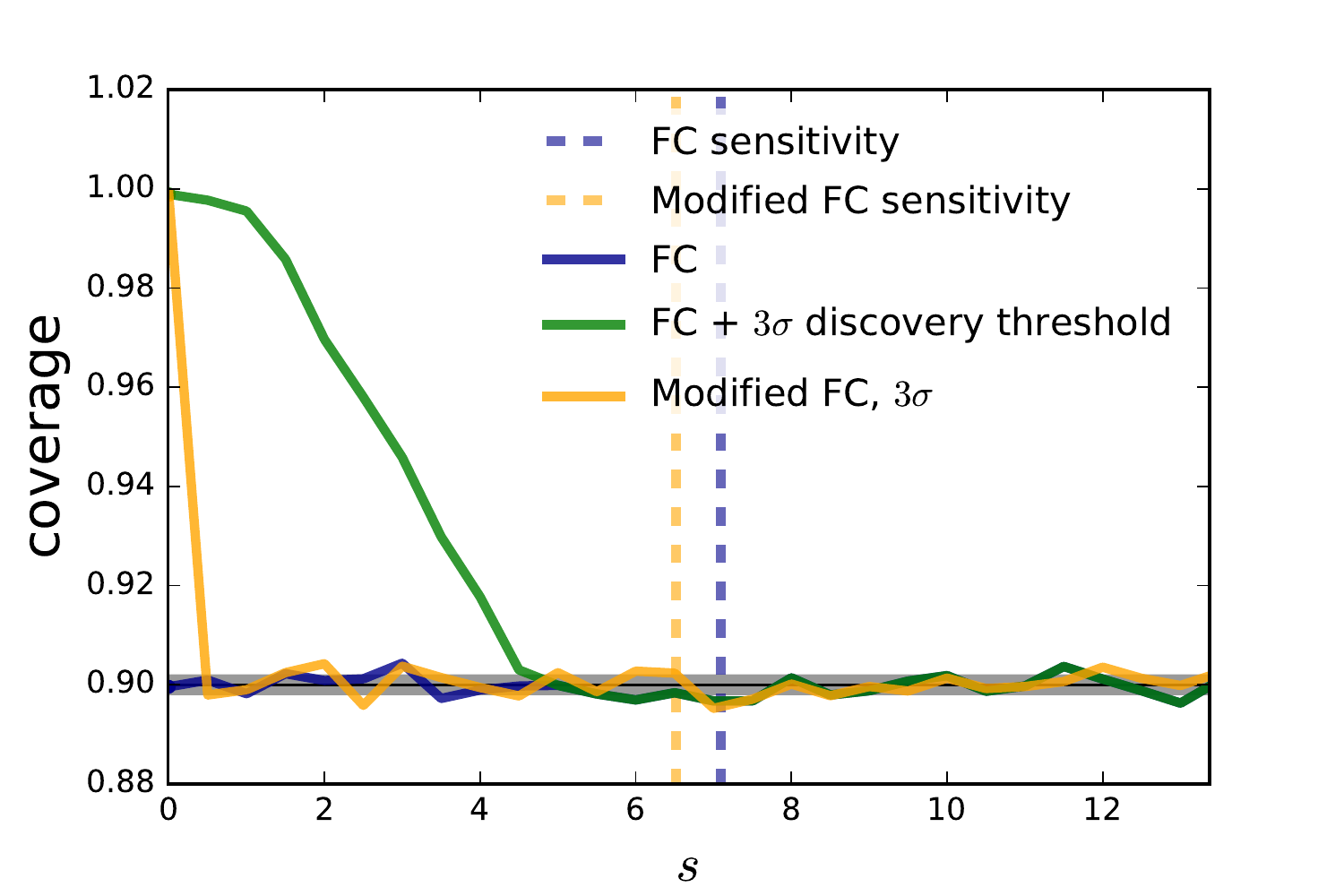}}
    \subfloat[$m=5$, $4\sigma$ threshold]{\includegraphics[width=0.5\linewidth]{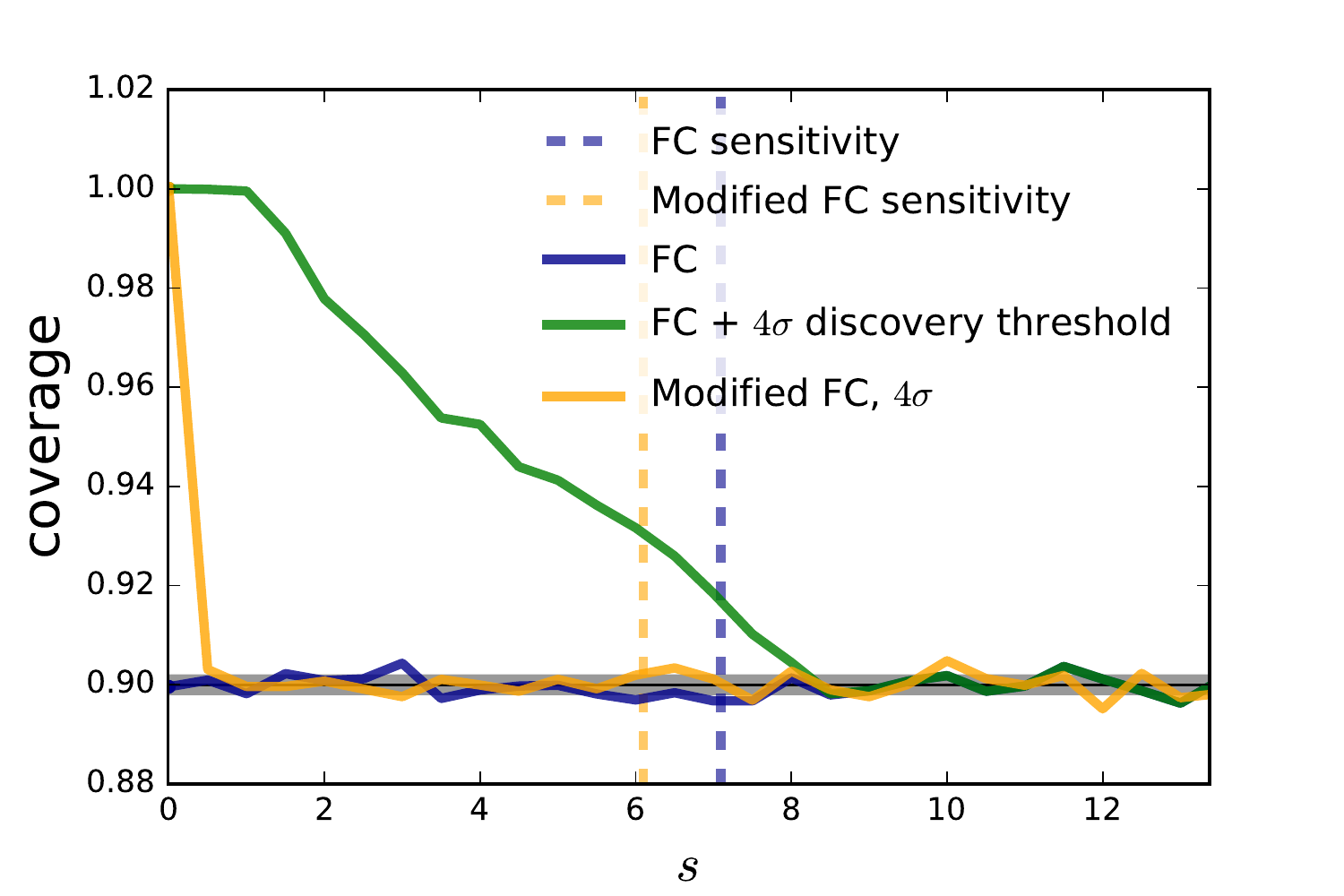}}\\
    \subfloat[$m=60$, $3\sigma$ threshold]{\includegraphics[width=0.5\linewidth]{powerlaw_coverage_mass_60_sigmad_3.pdf}}
    \subfloat[$m=60$, $4\sigma$ threshold]{\includegraphics[width=0.5\linewidth]{powerlaw_coverage_mass_60_sigmad_4.pdf}}
    \caption{Coverage as function of signal expectation for the FC method (blue), the FC method, including a $3$ or $4\sigma$ discovery threshold (green) and the modified FC method (orange), for two line masses. Median upper limits are indicated in dashed lines for the FC (blue) and modified FC (orange) cases.}
    \label{fig:coverages}
\end{figure}

The coverage of the FC and modified FC method $90\%$ confidence intervals are shown in fig~\ref{fig:coverages} for $m=5,60$. The pure FC method, shown in a blue line, provides the expected coverage. The green curves show the over-coverage of experiments using the FC construction with a threshold for reporting the lower limit of either a $3$ or $4\sigma$ discovery significance. The modified FC method exhibits the desired coverage of $0.9$ for signals greater than $0$, but jumps to cover the $0$-signal case with the desired coverage corresponding to the discovery threshold. A slight improvement in the median upper limits, indicated with dashed lines, for the modified FC method, due to the tighter lower threshold shown in figure~\ref{fig:ciconstruction} can be seen, with a greater change seen for the larger, $4\sigma$ discovery threshold. 

%

\section{Summary}
\label{sec:summary}
This paper proposes a method for constructing a modified FC method where the discovery significance is different from the confidence level of the upper limits and intervals. For an example case, the coverage at 0 signal corresponds to the discovery significance, and moves to the required confidence level for all signals larger than 0. This allows experiments to avoid the over-coverage that results from expanding the standard FC intervals, and simplifies reporting or discussion of coverage properties. The intervals approach the one-sided upper limit for under-fluctuations of the data, motivating an application of a power-constraint lower signal threshold or similar to the confidence interval construction outlined in this paper. This will result in discrete coverage regimes that depend on the true signal size, and allow the experiment to directly set the discovery threshold and minimal discovery power independent of the confidence level of the interval. 
%

\section*{Acknowledgements}
The author would like to thank Jan Conrad and Jelle Aalbers   for  fruitful discussions and suggestions.
This research was supported by a grant of the Knut and Alice Wallenberg Foundation, PI: J. Conrad

\bibliography{main.bib}

\end{document}